\documentclass[
    ,final            
  ]
  {aipproc}

\layoutstyle{8x11single}

\newcommand{\gray}[1]{$\gamma$-ray{#1}}

\newcommand{\pubjournal}[6] {#1, #2 {\bf #3}, #4 (#5).}

\newcommand{\aap}{{\it Astron. Astrophys.}}
\newcommand{\apj}{{\it ApJ}}
\newcommand{\prd}{{\it Phys. Rev. D}}
\newcommand{\prl}{{\it Phys. Rev. Lett.}}
\newcommand{\physrep}{{\it Phys. Rep.}}
\newcommand{\mnras}{{\it Mon. Not. Royal Astron. Soc.}}

\begin{document}

\title{Identifying Dark Matter Burners in the Galactic center}



\author{Igor V. Moskalenko}{
  address={Hansen Experimental Physics Laboratory,
  Stanford University, Stanford, CA 94305},
  altaddress={Kavli Institute for Particle Astrophysics and Cosmology,
    Stanford University, Stanford, CA 94309}
}

\author{Lawrence L. Wai}{
  address={Stanford Linear Accelerator Center,
  2575 Sand Hill Rd, Menlo Park, CA 94025},
  altaddress={Kavli Institute for Particle Astrophysics and Cosmology,
    Stanford University, Stanford, CA 94309}
}

\begin{abstract}
If the supermassive black hole (SMBH) at the center of our Galaxy grew
adiabatically, then a dense "spike" of dark matter is expected to have
formed around it. Assuming that dark matter is composed primarily of
weakly interacting massive particles (WIMPs), a star orbiting close
enough to the SMBH can capture WIMPs at an extremely high rate. The
stellar luminosity due to annihilation of captured WIMPs in the
stellar core may be comparable to or even exceed the luminosity of the
star due to thermonuclear burning. The model thus predicts the
existence of unusual stars, i.e. "WIMP burners", in the vicinity of an
adiabatically grown SMBH. We find that the most efficient WIMP burners
are stars with degenerate electron cores, e.g. white dwarfs (WD) or
degenerate cores with envelopes. If found, such stars would provide
evidence for the existence of particle dark matter and could possibly
be used to establish its density profile. In our previous paper we
computed the luminosity from WIMP burning for a range of dark matter
spike density profiles, degenerate core masses, and distances from the
SMBH. Here we compare our results with the observed stars closest to
the Galactic center and find that they could be consistent with WIMP
burners in the form of degenerate cores with envelopes. We also
cross-check the WIMP burner hypothesis with the EGRET observed flux of
gamma-rays from the Galactic center, which imposes a constraint on the
dark matter spike density profile and annihilation cross-section. We
find that the EGRET data is consistent with the WIMP burner
hypothesis. New high precision measurements by GLAST will confirm or
set stringent limits on a dark matter spike at the Galactic center,
which will in turn support or set stringent limits on the existence of
WIMP burners at the Galactic center.

\end{abstract}

\maketitle


\section{Results}

The highest density ``free space'' dark matter regions occur for dark
matter particles captured within the gravitational potential of
adiabatically grown SMBHs.
Any star close enough to 
such a
SMBH can capture a large number of WIMPs during a short period of
time.  Annihilation of captured WIMPs may lead to considerable energy
release in stellar cores thus affecting the evolution and appearance
of such stars.  Such an idea has been first proposed in
\cite{salati89} and further developed in \cite{bouquet89} who applied
it to main-sequence stars.  An order-of-magnitude estimate of the WIMP
capture rates for stars of various masses and evolution stages
\cite{MW2006} lead us to the conclusion that WDs, fully burned stars
without their own energy supply, are the most promising candidates to
look for. WIMP capture by WDs or degenerate cores with envelopes
located in a high density dark matter region has been discussed in
detail in \cite{MW2007}. 

A high WIMP concentration in the stellar interior may affect the
evolution and appearance of a star. The effects of WIMPs can be
numerous, here we list only a few.  The additional source of energy
from WIMP pair-annihilation may cause convective energy transport from
the stellar interior when radiative transport is not effective
enough. In turn, this may inflate the stellar radius.  On the other
hand, WIMPs themselves may provide energy transport and suppress
convection in the stellar core; this would reduce the replenishment of
the thermonuclear burning region with fresh fuel. 
The appearance of massive stars and the bare WDs should not
change, however. The former are too luminous, $L_*\propto M_*^4$,
while the energy transport in the latter is dominated by the
degenerate electrons.  Here we discuss observational features of DM
burners, and GLAST's role in checking this hypothesis.

There several possible ways to identify the DM burners:

\begin{itemize}

\item  The bare WDs burning DM should be hot, with luminosity maximum falling
into the UV or X-ray band.  The number of very hot WDs in the SDSS
catalog \cite{e06} is small, just a handful out of 9316. This means
that observation of a concentration of very hot WDs at the GC would be
extremely unlikely unless they are ``DM burners.''

\item Identification of DM burners may be possible by combining the data
obtained by several experiments:

  \begin{itemize}

  \item GLAST \gray{} measurements from the GC can be used to identify
  a putative DM spike at the SMBH, and also measure the annihilation
  flux from the spike.  Identification of the DM spike requires
  a detection of a point source at the GC (i.e. not extended)
  centered on the SMBH (i.e. with no offset), and a source
  spectrum matching a WIMP of a particular mass, which agrees with the
  ``universal'' WIMP mass as determined by any other putative WIMP
  signals (i.e. from colliders, direct detection, other indirect
  detection).

  \item Direct measurement of the WIMP-nucleon scattering
  cross-section fixes the WIMP capture rate and thus the WIMP burner
  luminosity for a given degenerate core.

  \item Determination of stellar orbits would allow a calculation of
  the WIMP burning rate by a particular star and, therefore, the
  proportion of its luminosity which is coming from the WIMP burning.

  \item LHC measurements may provide information about the WIMP mass
  and interaction cross-sections.

 \end{itemize}


\end{itemize}

\begin{figure}
  \includegraphics[height=.28\textheight]{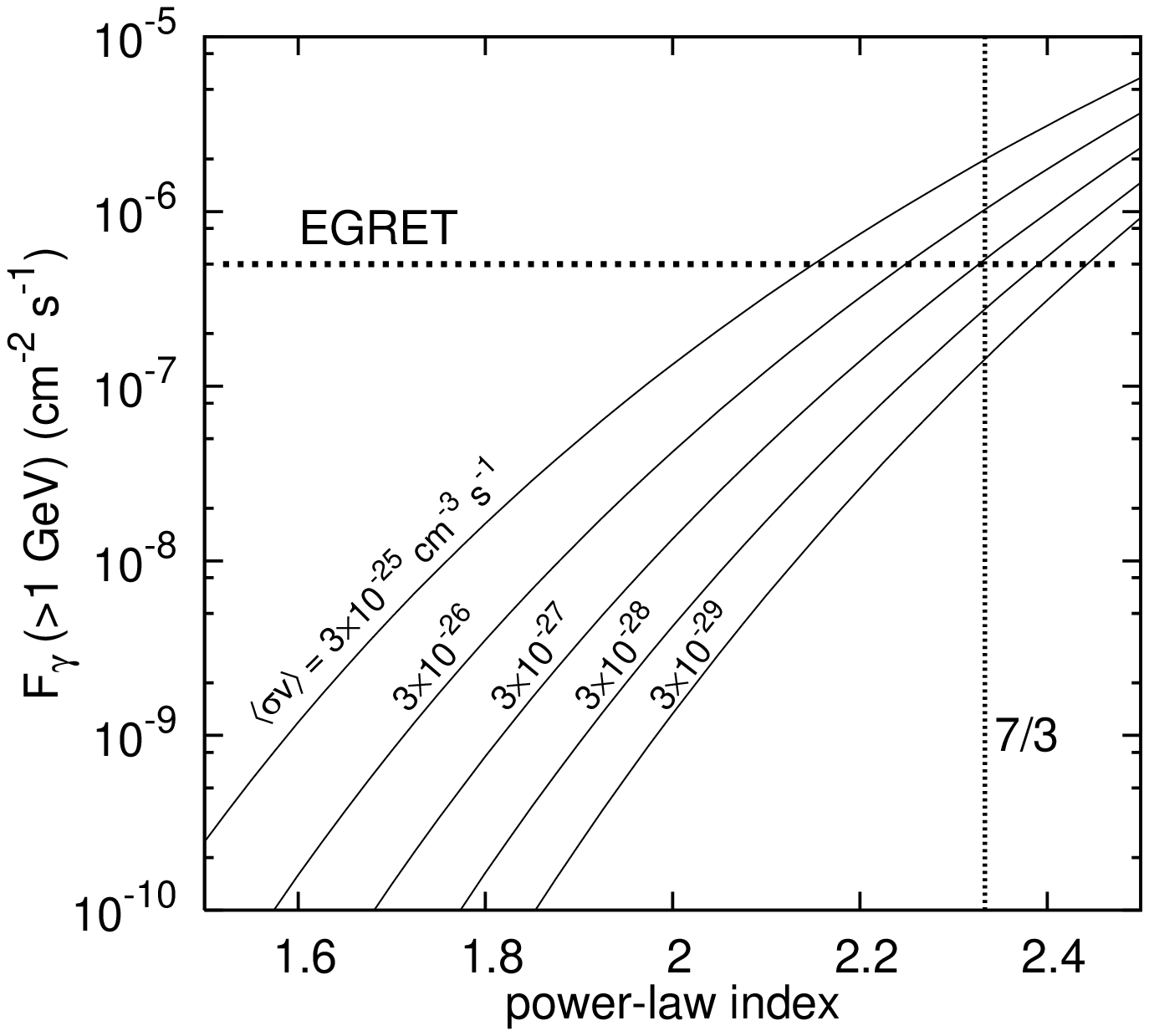}
  \includegraphics[height=.28\textheight]{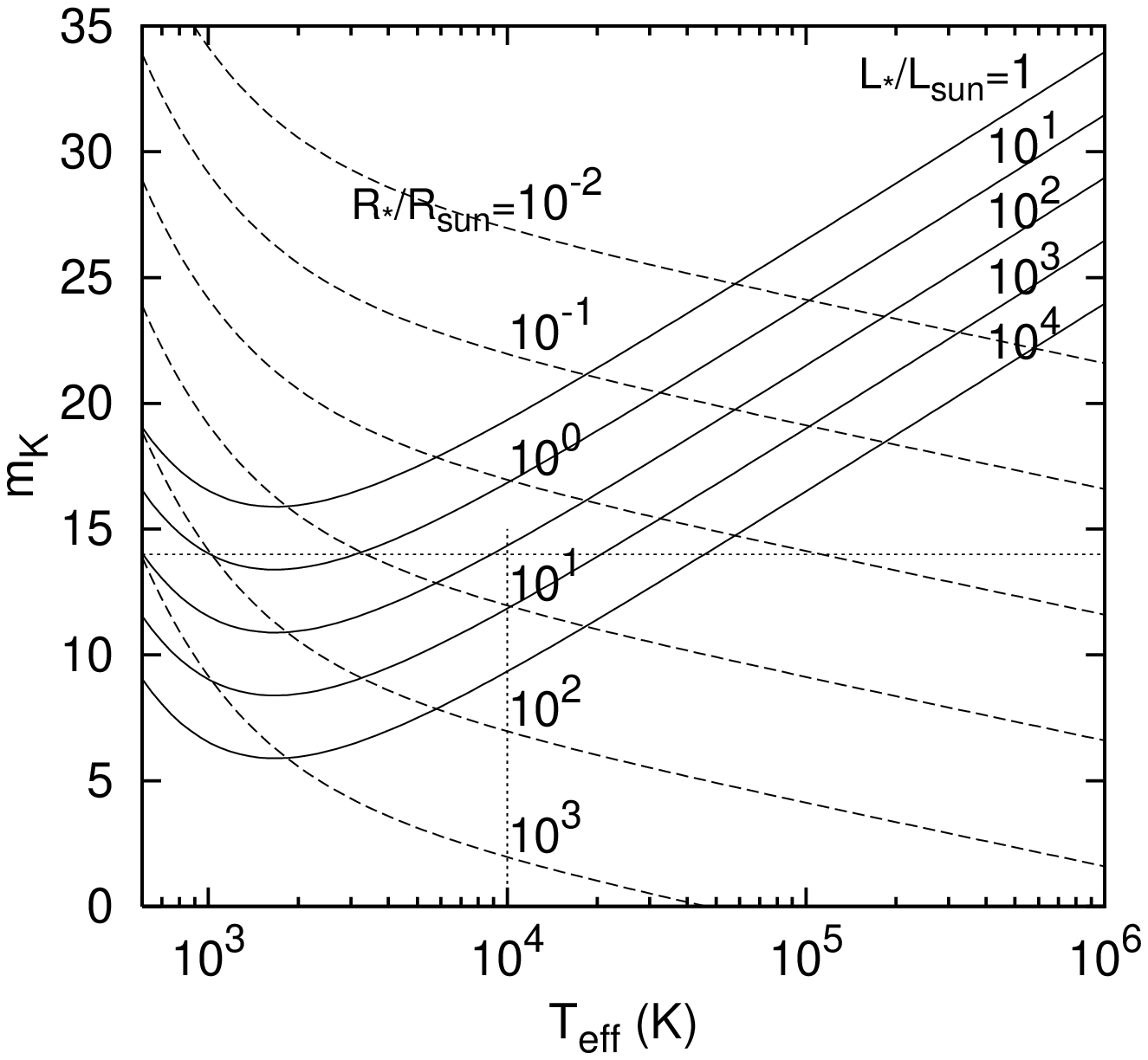}
  \caption{{\it Left:} \gray{} flux vs.\ the DM central spike
power-law index. The lines are shown for a series
of annihilation cross sections $\langle\sigma v\rangle$.
{\it Right:} The visual K-band magnitude of DM burners at the GC
without extinction vs.\ the effective surface temperature.
}
\end{figure}

Figure 1 (left) shows the DM annihilation \gray{} flux from the
central spike vs.\ DM density power-law index assuming 10$\gamma$'s
above 1 GeV per annihilation and WIMP mass $m_\chi=100$ GeV. The EGRET
\gray{} flux from the GC $F_\gamma(>1\ {\rm GeV})=5\times
10^{-7}$ cm$^{-2}$ s$^{-1}$ \cite{MH98}.

Advances in near-IR instrumentation have made possible observations of
stars in the inner parsec of the Galaxy
\citep{genzel00,ghez03,ghez05}. The apparent K-band brightness of
these stars is 14--17 mag while the extinction may be as large as 3.3
mag \citep{rrp89}. Assuming a central spike with index 7/3, the K-band
brightness for bare Oxygen WDs with $T_{\rm eff}\sim100,000$~K and
$R_*/R_\odot\sim0.01$ is about 22--23 mag not including extinction.  A
WIMP burning degenerate core with envelope may be cold enough to
produce most of its emission in the IR band (Figure 1, right).  For a
given luminosity, the colder stars should necessarily have larger
outer radii. A DM burner (w/envelope) with effective temperature
$T_{\rm eff}<10,000$ K and radius $>5R_\odot$ could have visual K-band
magnitude mag $>10$ (without extinction) and be visible with the
current techniques. The horizontal dotted line (mag = 14) show the
dimmest stars currently observed in the GC.


I.~V.~M.\ acknowledges partial support from a NASA APRA grant.  
A part of this work was done at Stanford Linear Accelerator Center, 
Stanford University, and supported by Department of Energy contract
DE-AC03-768SF00515.



\bibliographystyle{aipproc}   

\begin{thebibliography}{9}

\bibitem{salati89} 
\pubjournal{P. Salati, and J. Silk}{\apj}{338}{24--31}{1989}{}

\bibitem{bouquet89} 
\pubjournal{A. Bouquet, and  P. Salati}{\apj}{346}{284--288}{1989}{}

\bibitem{MW2006}
I.~V. Moskalenko, and L.~L. Wai, \emph{arXiv:} astro-ph/0608535 (2006).

\bibitem{MW2007}
\pubjournal{I.~V. Moskalenko, and L.~L. Wai}{\apj}{659}{L29--L32}{2007}{}

\bibitem{e06} 
\pubjournal{D.~J. Eisenstein et al.}{\apj S}{167}{40--58}{2006}{}

\bibitem{MH98}
\pubjournal{H.~A. Mayer-Hasselwander et al.}{\aap}{335}{161--172}{1998}

\bibitem{genzel00} 
\pubjournal{R. Genzel et al.}{\mnras}{317}{348--374}{2000}
{Stellar dynamics in the Galactic Centre: proper motions and anisotropy}

\bibitem{ghez03} 
\pubjournal{A.~M. Ghez et al.}{\apj}{586}{L127--L131}{2003}{}

\bibitem{ghez05} 
\pubjournal{A.~M. Ghez et al.}{\apj}{620}{744--757}{2005}{}

\bibitem{rrp89} 
\pubjournal{G.~H. Rieke, M.~J. Rieke, and A.~E. Paul}{\apj}{336}{752--761}{1989}{}

\end{thebibliography}

\end{document}